\documentclass[12pt]{article}
\usepackage{setspace}
\usepackage{amsfonts, epsfig}
\begin{document}
\title{
  \vskip 15pt {\bf A comment on \lq{}A fast $\mathcal{L}_p$ spike alignment metric\rq{} by A.~J. Dubbs, B.~A. Seiler and M.~O. Magnasco [arXiv:0907.3137]} \author{
    {\large Conor Houghton\footnote{houghton@maths.tcd.ie}}\\ 
{\normalsize {\sl School of Mathematics, Trinity College Dublin, Ireland}}
}
\date{27 July 2009}}

\maketitle

\begin{abstract}
Measuring the transmitted information in metric-based clustering has
become something of a standard test for the performance of a spike
train metric. In this comment, the recently proposed $\mathcal{L}_p$
Victor-Purpura metric is used to cluster spiking responses to zebra
finch songs, recorded from field L of anesthetized zebra finch. It is
found that for these data the $\mathcal{L}_p$ metrics with $p>1$
modestly outperform the standard, $p=1$, Victor-Purpura metric. It is
argued that this is because for larger values of $p$, the metric comes
closer to performing windowed coincidence detection.
\end{abstract}

\section{Introduction}

There are a number of reasons why metrics are believed to be useful
for studying spike trains. It can be argued that it is the most
general, useful, mathematical framework for spike trains
\cite{VictorPurpura1997a}, it may be possible to find a manifold of
spike trains using local linear methods and, of most relevance to the
present discussion, it is possible that studying spike train metrics
is a useful approach to understanding how content is coded in spike
trains.

The Victor-Purpura metric is an edit distance metric on the space of
spike trains. The distance between two spike trains is, in effect,
calculated as the cost of changing one spike train into the other by
adding, deleting, or moving spikes, with an individual cost for each
type of elementary moves. In particular, a cost of one is associated
with adding or deleting a spike and a cost of $q\delta t$ with moving
a spike a time $\delta t$: the distance is
\begin{equation}
d({\bf u},{\bf v};q)=\mbox{min}_\gamma c_{\gamma;q}({\bf u},{\bf v})
\end{equation}
where $c_\gamma({\bf u},{\bf v})$ is the cost of a sequence of
elementary moves $\gamma$ and is calculated by adding the cost of all
the elementary moves in the sequence. The minimum is taken over all
sequences $\gamma$ changing ${\bf u}$ to ${\bf v}$. The parameter
$2/q$ gives an important timescale for the metric: it is never
worthwhile to move a spike more than $2/q$ since it would be cheaper
to delete the spike from one temporal location and to add it to the
other. This gives a timescale that, roughly speaking, separates jitter
from unreliability in comparing spikes in the spike trains. However,
the Victor-Purpura metric does more than this; it explicitly pairs up
those spikes that can be thought of as being related by jitter.

The Victor-Purpura metric has an $l^1$ character: the cost $c_\gamma$
is a simple linear sum of the individual costs of the individual
moves. The generalization proposed in \cite{DubbsSeilerMagnasco2009a}
changes this to an $l^p$-like sum. For notational convenience, from
now on the set of sequences will be restricted: the minimum sequence
will never involve moving a spike after it has been added, it would
always be cheaper to add the spike in the correct location, similarly,
spikes are never moved before they are deleted. It is also specified
that spikes can only be moved once. With these restrictions, the
sequence $\gamma$ can be considered to be an unordered set made up of
deletions, additions and moves. Let
\begin{equation}
\gamma=\alpha\cup\mu
\end{equation}
where $\alpha$ is the set of additions and deletions and $\mu$ is the
set of moves. The elements of $\mu$ are pairs of spikes $(u,v)$, with
$u\in {\bf u}$ and $v\in {\bf v}$, that are related by jitter.

Now,
\begin{equation}
c_{\gamma;q}({\bf u},{\bf v})=|\alpha|+\sum_{(u,v)\in\mu} q|u-v|
\end{equation}
This is generalized in \cite{DubbsSeilerMagnasco2009a} to
\begin{equation}
c_{\gamma;q,p}({\bf u},{\bf v})=\left(|\alpha|+\sum_{(u,v)\in\mu} q^p|u-v|^p\right)^{1/p}
\end{equation}
with $p\ge 1$, to give the $\mathcal{L}_p$ Victor-Purpura metric
\begin{equation}
d({\bf u},{\bf v};q,p)=\mbox{min}_\gamma c_{\gamma;q,p}({\bf u},{\bf v})
\end{equation}
where the minimum is taken over sequences $\gamma$ changing ${\bf u}$
to ${\bf v}$ and satisfying the restrictions specified above. An
algorithm for calculating this quantity is given in
\cite{DubbsSeilerMagnasco2009a}; the existing algorithms used to
calculate the Victor-Purpura metric,
\cite{VictorGoldbergGardner2007a}, for example, could also be adapted
to $p>1$.

\section{Evaluating the metric using clustering}

In the data that will be considered here, spike trains were recorded
from field L of the auditory fore-brain of anesthetized zebra finch
during playback of 20 con-specific songs with each song repeated ten
times, to give a total of 200 spike trains. These spike trains, and
the experimental conditions used to produce them, are described in
\cite{NarayanEtAl2006b,WangEtAl2007a} and they have previously been
used to compare metrics in
\cite{Houghton2007a,HoughtonVictor2009a}. The key point, in this
context, is that a good metric, one which depends on the content
encoded in the spike train, should measure a smaller distance between
spiking responses to the same song than between responses to different
songs. One motivation for looking at spike train metrics is that
scoring metrics in this way, and examining how good metrics measure
distance, might reveal details about how content is encoded.

The usual method for evaluating how well a metric succeeds in this way
is to calculate the transmitted information
\cite{VictorPurpura1996a}. There are two ways to cluster the spike
trains: a true clustering, based on the identity of the song that
elicited the response and an estimated clustering, based on the metric
distances between the responses. Roughly speaking, the transmitted
information quantifies the amount of information about the true
clustering given by the estimated clustering.

Calculating the transmitted information relies on the calculation of a
confusion matrix $N$. This is described in detail in
\cite{VictorPurpura1996a}, described specifically for the data
considered here, $N$ is a $20\times 20$ matrix of integers where the
entry $N_{ij}$ counts how many of the spike trains from the $i$th true
cluster is closest, on average, to the spike trains in the $j$th true
cluster. As in \cite{VictorPurpura1996a} this averaging is carried out
as a root-mean-square averaging to under-weigh outliers. If the metric
clustering is close to the true clustering, $N$ will be nearly
diagonal. The transmitted information gives a measure of this, it is
\begin{equation}
\tilde{h}=\frac{1}{n\ln{20}}\sum_{ij}N_{ij}\left(\ln{N_{ij}}-\ln{\sum_k{N_{kj}}}-\ln{\sum_k{N_{ik}}}+\ln {n}\right).
\end{equation}
where $n=\sum_i\sum_jN_{ij}$. All sums are from one to 20: the number
of clusters. The factor of $\ln{20}$ normalizes $\tilde{h}$ and so
that it takes values between zero and one; a low value corresponds to
poor clustering and a value near one corresponds to a near-diagonal
confusion matrix.

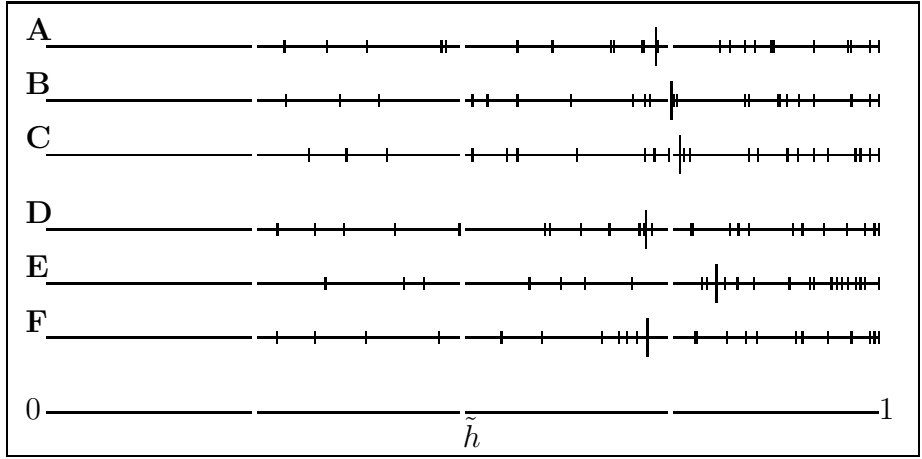
\begin{figure}

\begin{center}
\setlength{\unitlength}{0.01575pt}
\framebox{
\begin{picture}(20999,10500)(0,0)
\put(0,700){{0}}
\put(10499,25){{$\tilde{h}$}}
\put(20499,700){{1}}
\put(500,850){\line(1,0){4924}}
\put(5574,850){\line(1,0){4849}}
\put(10574,850){\line(1,0){4849}}
\put(15574,850){\line(1,0){4924}}
\put(6030,2500){\line(0,1){300}}
\put(6945,2500){\line(0,1){300}}
\put(8182,2500){\line(0,1){300}}
\put(9926,2500){\line(0,1){300}}
\put(9934,2500){\line(0,1){300}}
\put(11429,2500){\line(0,1){300}}
\put(12395,2500){\line(0,1){300}}
\put(13853,2500){\line(0,1){300}}
\put(14254,2500){\line(0,1){300}}
\put(14439,2500){\line(0,1){300}}
\put(14686,2500){\line(0,1){300}}
\put(14946,2500){\line(0,1){300}}
\put(16081,2500){\line(0,1){300}}
\put(16128,2500){\line(0,1){300}}
\put(16837,2500){\line(0,1){300}}
\put(17302,2500){\line(0,1){300}}
\put(17577,2500){\line(0,1){300}}
\put(18511,2500){\line(0,1){300}}
\put(18663,2500){\line(0,1){300}}
\put(19278,2500){\line(0,1){300}}
\put(19838,2500){\line(0,1){300}}
\put(20274,2500){\line(0,1){300}}
\put(20388,2500){\line(0,1){300}}
\put(20496,2500){\line(0,1){300}}
\put(0,2800){{\bf F}}
\put(500,2650){\line(1,0){4924}}
\put(5574,2650){\line(1,0){4849}}
\put(10574,2650){\line(1,0){4849}}
\put(15574,2650){\line(1,0){4924}}
\put(14933,2650){\line(0,1){450}}
\put(14933,2650){\line(0,-1){450}}
\put(7203,3800){\line(0,1){300}}
\put(9090,3800){\line(0,1){300}}
\put(9568,3800){\line(0,1){300}}
\put(12104,3800){\line(0,1){300}}
\put(12860,3800){\line(0,1){300}}
\put(13432,3800){\line(0,1){300}}
\put(14572,3800){\line(0,1){300}}
\put(16247,3800){\line(0,1){300}}
\put(16365,3800){\line(0,1){300}}
\put(16796,3800){\line(0,1){300}}
\put(17102,3800){\line(0,1){300}}
\put(17495,3800){\line(0,1){300}}
\put(18349,3800){\line(0,1){300}}
\put(18843,3800){\line(0,1){300}}
\put(18945,3800){\line(0,1){300}}
\put(19353,3800){\line(0,1){300}}
\put(19497,3800){\line(0,1){300}}
\put(19599,3800){\line(0,1){300}}
\put(19763,3800){\line(0,1){300}}
\put(19947,3800){\line(0,1){300}}
\put(20052,3800){\line(0,1){300}}
\put(20161,3800){\line(0,1){300}}
\put(20496,3800){\line(0,1){300}}
\put(20499,3800){\line(0,1){300}}
\put(0,4100){{\bf E}}
\put(500,3950){\line(1,0){4924}}
\put(5574,3950){\line(1,0){4849}}
\put(10574,3950){\line(1,0){4849}}
\put(15574,3950){\line(1,0){4924}}
\put(16597,3950){\line(0,1){450}}
\put(16597,3950){\line(0,-1){450}}
\put(6049,5100){\line(0,1){300}}
\put(6951,5100){\line(0,1){300}}
\put(7649,5100){\line(0,1){300}}
\put(8878,5100){\line(0,1){300}}
\put(10424,5100){\line(0,1){300}}
\put(12478,5100){\line(0,1){300}}
\put(12600,5100){\line(0,1){300}}
\put(13347,5100){\line(0,1){300}}
\put(14025,5100){\line(0,1){300}}
\put(14745,5100){\line(0,1){300}}
\put(14857,5100){\line(0,1){300}}
\put(15047,5100){\line(0,1){300}}
\put(15980,5100){\line(0,1){300}}
\put(16024,5100){\line(0,1){300}}
\put(16921,5100){\line(0,1){300}}
\put(17125,5100){\line(0,1){300}}
\put(17379,5100){\line(0,1){300}}
\put(18440,5100){\line(0,1){300}}
\put(18663,5100){\line(0,1){300}}
\put(19170,5100){\line(0,1){300}}
\put(19730,5100){\line(0,1){300}}
\put(20161,5100){\line(0,1){300}}
\put(20388,5100){\line(0,1){300}}
\put(20496,5100){\line(0,1){300}}
\put(0,5400){{\bf D}}
\put(500,5250){\line(1,0){4924}}
\put(5574,5250){\line(1,0){4849}}
\put(10574,5250){\line(1,0){4849}}
\put(15574,5250){\line(1,0){4924}}
\put(14897,5250){\line(0,1){450}}
\put(14897,5250){\line(0,-1){450}}
\put(6810,6900){\line(0,1){300}}
\put(7705,6900){\line(0,1){300}}
\put(8680,6900){\line(0,1){300}}
\put(10738,6900){\line(0,1){300}}
\put(11557,6900){\line(0,1){300}}
\put(11812,6900){\line(0,1){300}}
\put(13248,6900){\line(0,1){300}}
\put(14872,6900){\line(0,1){300}}
\put(15103,6900){\line(0,1){300}}
\put(15458,6900){\line(0,1){300}}
\put(15810,6900){\line(0,1){300}}
\put(15952,6900){\line(0,1){300}}
\put(17369,6900){\line(0,1){300}}
\put(17589,6900){\line(0,1){300}}
\put(18284,6900){\line(0,1){300}}
\put(18313,6900){\line(0,1){300}}
\put(18547,6900){\line(0,1){300}}
\put(18935,6900){\line(0,1){300}}
\put(19267,6900){\line(0,1){300}}
\put(19931,6900){\line(0,1){300}}
\put(20055,6900){\line(0,1){300}}
\put(20274,6900){\line(0,1){300}}
\put(20496,6900){\line(0,1){300}}
\put(20499,6900){\line(0,1){300}}
\put(0,7200){{\bf C}}
\put(500,7050){\line(1,0){4924}}
\put(5574,7050){\line(1,0){4849}}
\put(10574,7050){\line(1,0){4849}}
\put(15574,7050){\line(1,0){4924}}
\put(15721,7050){\line(0,1){450}}
\put(15721,7050){\line(0,-1){450}}
\put(6248,8200){\line(0,1){300}}
\put(7545,8200){\line(0,1){300}}
\put(8476,8200){\line(0,1){300}}
\put(10738,8200){\line(0,1){300}}
\put(11095,8200){\line(0,1){300}}
\put(11812,8200){\line(0,1){300}}
\put(13105,8200){\line(0,1){300}}
\put(14587,8200){\line(0,1){300}}
\put(14885,8200){\line(0,1){300}}
\put(15004,8200){\line(0,1){300}}
\put(15578,8200){\line(0,1){300}}
\put(15651,8200){\line(0,1){300}}
\put(17276,8200){\line(0,1){300}}
\put(17369,8200){\line(0,1){300}}
\put(18090,8200){\line(0,1){300}}
\put(18120,8200){\line(0,1){300}}
\put(18293,8200){\line(0,1){300}}
\put(18568,8200){\line(0,1){300}}
\put(18935,8200){\line(0,1){300}}
\put(19838,8200){\line(0,1){300}}
\put(19857,8200){\line(0,1){300}}
\put(20274,8200){\line(0,1){300}}
\put(20496,8200){\line(0,1){300}}
\put(20499,8200){\line(0,1){300}}
\put(0,8500){{\bf B}}
\put(500,8350){\line(1,0){4924}}
\put(5574,8350){\line(1,0){4849}}
\put(10574,8350){\line(1,0){4849}}
\put(15574,8350){\line(1,0){4924}}
\put(15514,8350){\line(0,1){450}}
\put(15514,8350){\line(0,-1){450}}
\put(6218,9500){\line(0,1){300}}
\put(7242,9500){\line(0,1){300}}
\put(8200,9500){\line(0,1){300}}
\put(9985,9500){\line(0,1){300}}
\put(10102,9500){\line(0,1){300}}
\put(11812,9500){\line(0,1){300}}
\put(12659,9500){\line(0,1){300}}
\put(14064,9500){\line(0,1){300}}
\put(14141,9500){\line(0,1){300}}
\put(14812,9500){\line(0,1){300}}
\put(14850,9500){\line(0,1){300}}
\put(15196,9500){\line(0,1){300}}
\put(16686,9500){\line(0,1){300}}
\put(16925,9500){\line(0,1){300}}
\put(17279,9500){\line(0,1){300}}
\put(17515,9500){\line(0,1){300}}
\put(17908,9500){\line(0,1){300}}
\put(17965,9500){\line(0,1){300}}
\put(18935,9500){\line(0,1){300}}
\put(19745,9500){\line(0,1){300}}
\put(19832,9500){\line(0,1){300}}
\put(20274,9500){\line(0,1){300}}
\put(20496,9500){\line(0,1){300}}
\put(20499,9500){\line(0,1){300}}
\put(0,9800){{\bf A}}
\put(500,9650){\line(1,0){4924}}
\put(5574,9650){\line(1,0){4849}}
\put(10574,9650){\line(1,0){4849}}
\put(15574,9650){\line(1,0){4924}}
\put(15139,9650){\line(0,1){450}}
\put(15139,9650){\line(0,-1){450}}
\end{picture}
}
\end{center}

\caption{Performance as $p$ changes. {\bf A}, {\bf B} and {\bf C}
  correspond to $p=1$, $p=2$ and $p=10$. Each horizontal line runs
  from zero to one. For clarity, there are tiny gaps at 0.25, 0.5 and
  0.75. Each small vertical dash marks the $\tilde{h}$ value for one
  site where for each site the optimal value of $q$ is used. The large
  vertical dash marks the average, 0.732, 0.751 and 0.761
  respectively. To indicate the range over which $\tilde{h}$ values
  vary, two other metrics are plotted for comparison. ${\bf D}$
  corresponds to the van Rossum metric \cite{vanRossum2001a}, with
  each site using its optimal value of $\tau$. ${\bf E}$ corresponds
  to the synapse metric, with each site using its optimal value of
  $\tau$ and $\mu$ \cite{Houghton2007a}. Finally, ${\bf F}$
  corresponds to the $L^1$ van Rossum metric, with a boxcar filter;
  this metric measures a very similar distance to the $p=1$ metric
  \cite{HoughtonSen2006a}.\label{Results}}
\end{figure}

The main purpose of this comment is to evaluate the $\mathcal{L}_p$
Victor-Purpura metric by calculating $\tilde{h}$ for the zebra finch
data previous used to evaluate other spike train metrics
\cite{Houghton2007a,HoughtonVictor2009a}. The data set contains 24
sets of spike trains, corresponding to 24 recording sites from
multiple birds. Although the songs themselves are of different
lengths, all of the songs are at least a second long and, in each
case, the spike train is truncated to one second, starting at song
onset.

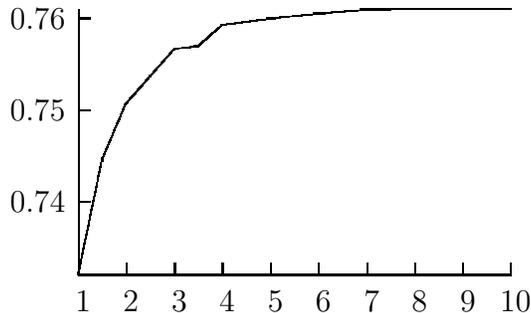
\begin{figure}
\begin{center}

\setlength{\unitlength}{0.240900pt}
\ifx\plotpoint\undefined\newsavebox{\plotpoint}\fi
\sbox{\plotpoint}{\rule[-0.200pt]{0.400pt}{0.400pt}}%
\begin{picture}(900,540)(0,0)
\sbox{\plotpoint}{\rule[-0.200pt]{0.400pt}{0.400pt}}%
\put(140,197){\makebox(0,0)[r]{ 0.74}}
\put(160.0,197.0){\rule[-0.200pt]{4.818pt}{0.400pt}}
\put(140,341){\makebox(0,0)[r]{ 0.75}}
\put(160.0,341.0){\rule[-0.200pt]{4.818pt}{0.400pt}}
\put(140,485){\makebox(0,0)[r]{ 0.76}}
\put(160.0,485.0){\rule[-0.200pt]{4.818pt}{0.400pt}}
\put(160,41){\makebox(0,0){ 1}}
\put(160.0,82.0){\rule[-0.200pt]{0.400pt}{4.818pt}}
\put(235,41){\makebox(0,0){ 2}}
\put(235.0,82.0){\rule[-0.200pt]{0.400pt}{4.818pt}}
\put(311,41){\makebox(0,0){ 3}}
\put(311.0,82.0){\rule[-0.200pt]{0.400pt}{4.818pt}}
\put(386,41){\makebox(0,0){ 4}}
\put(386.0,82.0){\rule[-0.200pt]{0.400pt}{4.818pt}}
\put(462,41){\makebox(0,0){ 5}}
\put(462.0,82.0){\rule[-0.200pt]{0.400pt}{4.818pt}}
\put(537,41){\makebox(0,0){ 6}}
\put(537.0,82.0){\rule[-0.200pt]{0.400pt}{4.818pt}}
\put(613,41){\makebox(0,0){ 7}}
\put(613.0,82.0){\rule[-0.200pt]{0.400pt}{4.818pt}}
\put(688,41){\makebox(0,0){ 8}}
\put(688.0,82.0){\rule[-0.200pt]{0.400pt}{4.818pt}}
\put(764,41){\makebox(0,0){ 9}}
\put(764.0,82.0){\rule[-0.200pt]{0.400pt}{4.818pt}}
\put(839,41){\makebox(0,0){ 10}}
\put(839.0,82.0){\rule[-0.200pt]{0.400pt}{4.818pt}}
\put(160.0,82.0){\rule[-0.200pt]{0.400pt}{100.696pt}}
\put(160.0,82.0){\rule[-0.200pt]{163.571pt}{0.400pt}}
\put(160,82){\usebox{\plotpoint}}
\multiput(160.58,82.00)(0.498,2.412){73}{\rule{0.120pt}{2.016pt}}
\multiput(159.17,82.00)(38.000,177.816){2}{\rule{0.400pt}{1.008pt}}
\multiput(198.58,264.00)(0.498,1.181){71}{\rule{0.120pt}{1.041pt}}
\multiput(197.17,264.00)(37.000,84.840){2}{\rule{0.400pt}{0.520pt}}
\multiput(235.58,351.00)(0.498,0.565){73}{\rule{0.120pt}{0.553pt}}
\multiput(234.17,351.00)(38.000,41.853){2}{\rule{0.400pt}{0.276pt}}
\multiput(273.58,394.00)(0.498,0.565){73}{\rule{0.120pt}{0.553pt}}
\multiput(272.17,394.00)(38.000,41.853){2}{\rule{0.400pt}{0.276pt}}
\multiput(311.00,437.59)(4.161,0.477){7}{\rule{3.140pt}{0.115pt}}
\multiput(311.00,436.17)(31.483,5.000){2}{\rule{1.570pt}{0.400pt}}
\multiput(349.00,442.58)(0.560,0.497){63}{\rule{0.548pt}{0.120pt}}
\multiput(349.00,441.17)(35.862,33.000){2}{\rule{0.274pt}{0.400pt}}
\multiput(386.00,475.58)(3.935,0.491){17}{\rule{3.140pt}{0.118pt}}
\multiput(386.00,474.17)(69.483,10.000){2}{\rule{1.570pt}{0.400pt}}
\multiput(462.00,485.59)(4.918,0.488){13}{\rule{3.850pt}{0.117pt}}
\multiput(462.00,484.17)(67.009,8.000){2}{\rule{1.925pt}{0.400pt}}
\multiput(537.00,493.59)(6.819,0.482){9}{\rule{5.167pt}{0.116pt}}
\multiput(537.00,492.17)(65.276,6.000){2}{\rule{2.583pt}{0.400pt}}
\put(613,498.67){\rule{18.067pt}{0.400pt}}
\multiput(613.00,498.17)(37.500,1.000){2}{\rule{9.034pt}{0.400pt}}
\put(688.0,500.0){\rule[-0.200pt]{36.376pt}{0.400pt}}
\put(160.0,82.0){\rule[-0.200pt]{0.400pt}{100.696pt}}
\put(160.0,82.0){\rule[-0.200pt]{163.571pt}{0.400pt}}
\end{picture}

\end{center}
\caption{The average performance plotted against $p$.
  $\tilde{h}$ has been calculated for a range of $p$ values using the
  optimal value of $q$ for each site and the average value of
  $\tilde{h}$ is plotted against $p$.\label{p}}
\end{figure}

The performance of the $\mathcal{L}_p$ Victor-Purpura metric is
illustrated in Fig.~\ref{Results}. A remarkable feature is that
$\tilde{h}$ rarely decreases as $p$ is increased. Other
multi-parameter metrics show a mixed effect; although changing a
parameter might increase the average $\tilde{h}$ across the sites,
there will typically be many individual sites where $\tilde{h}$
decreases. In the case of $p$, while the average improvement as $p$
increases is modest and a small number of sites show no change; there
is no site where the clustering is negatively affected when $p$ is
changed from one to two, three do show a tiny decrease between $p=1$
and $p=1.5$. $\tilde{h}$ increases for 19 sites between $p=1$ and
$p=2$ and for 15 between $p=2$ and $p=3$, the improvement gets less
and less and appears to have plateaued by $p=10$. The average
performance is plotted against $p$ in Fig.~{p}. The optimal value of
$q$ does not vary significantly as $p$ changes; however, the range of
$q$ that produces the optimal performance does get wider.

\section{Discussion}

It is interesting to compare this metric to another generalization of
the Victor-Purpura metric
\cite{VictorPurpura1996a,VictorPurpura1997a}. The linear distance
function $q\delta t$ can be replaced by any convex function,
$f(|\delta t|)$, of distance:
\begin{equation}
c_{\gamma;f(\cdot)}({\bf u},{\bf v})=|\alpha|+\sum_{(u,v)\in\mu} f(|u-v|).
\end{equation}
The distance function $f(|\delta t|)$ must be convex to ensure that
the triangular inequality is satisfied. Now, the function $q^p\delta
t^p$ in $c_{\gamma;q,p}$ above is concave for $p>1$. However, because
of the $1/p$ exponent in $d({\bf u},{\bf v};q,p)$, this does not cause
the same difficulty with the triangular inequality: an interesting
property of the ${\mathcal L}_p$ metrics is that they satisfy the
triangular inequality despite having a concave distance function.

As $p$ is increased, $q^p|\delta t|^p$ becomes increasingly concave
and spikes located at similar times contribute less and less to the
total cost and this cost is predominately made up of the cost of
adding and deleting spikes that are not related by jitter. This means
that the metric performs more and more as a windowed coincidence
detector as $p$ increases. The fact that it is always beneficial to
increase $p$ appears to imply that, for these data, coincidence and
coincidence detection has a role in the encoding and decoding of
content in spike trains.

\begin{figure}
\begin{center}
\setlength{\unitlength}{0.24pt}
\ifx\plotpoint\undefined\newsavebox{\plotpoint}\fi
\begin{picture}(1200,720)(0,0)
\sbox{\plotpoint}{\rule[-0.200pt]{0.400pt}{0.400pt}}%
\put(330,232){\makebox(0,0)[r]{ 0.25}}
\put(350.0,232.0){\rule[-0.200pt]{4.818pt}{0.400pt}}
\put(330,381){\makebox(0,0)[r]{ 0.5}}
\put(350.0,381.0){\rule[-0.200pt]{4.818pt}{0.400pt}}
\put(330,531){\makebox(0,0)[r]{ 0.75}}
\put(350.0,531.0){\rule[-0.200pt]{4.818pt}{0.400pt}}
\put(330,680){\makebox(0,0)[r]{ 1}}
\put(350.0,680.0){\rule[-0.200pt]{4.818pt}{0.400pt}}
\put(500,41){\makebox(0,0){ 0.25}}
\put(500.0,82.0){\rule[-0.200pt]{0.400pt}{4.818pt}}
\put(649,41){\makebox(0,0){ 0.5}}
\put(649.0,82.0){\rule[-0.200pt]{0.400pt}{4.818pt}}
\put(799,41){\makebox(0,0){ 0.75}}
\put(799.0,82.0){\rule[-0.200pt]{0.400pt}{4.818pt}}
\put(948,41){\makebox(0,0){ 1}}
\put(948.0,82.0){\rule[-0.200pt]{0.400pt}{4.818pt}}
\put(350.0,82.0){\rule[-0.200pt]{0.400pt}{144.058pt}}
\put(350.0,82.0){\rule[-0.200pt]{144.058pt}{0.400pt}}

\put(889,633){\line(1,0){8}}
\put(816,566){\line(1,0){8}}
\put(911,657){\line(1,0){8}}
\put(632,366){\line(1,0){8}}
\put(945,680){\line(1,0){8}}
\put(817,573){\line(1,0){8}}
\put(893,604){\line(1,0){8}}
\put(838,603){\line(1,0){8}}
\put(774,511){\line(1,0){8}}
\put(782,510){\line(1,0){8}}
\put(861,584){\line(1,0){8}}
\put(632,369){\line(1,0){8}}
\put(749,488){\line(1,0){8}}
\put(580,312){\line(1,0){8}}
\put(515,253){\line(1,0){8}}
\put(852,591){\line(1,0){8}}
\put(761,521){\line(1,0){8}}
\put(948,680){\line(1,0){8}}
\put(706,420){\line(1,0){8}}
\put(941,673){\line(1,0){8}}
\put(928,660){\line(1,0){8}}
\put(543,284){\line(1,0){8}}
\put(767,490){\line(1,0){8}}
\put(677,446){\line(1,0){8}}

\put(889,633){\line(-1,0){8}}
\put(816,566){\line(-1,0){8}}
\put(911,657){\line(-1,0){8}}
\put(632,366){\line(-1,0){8}}
\put(945,680){\line(-1,0){8}}
\put(817,573){\line(-1,0){8}}
\put(893,604){\line(-1,0){8}}
\put(838,603){\line(-1,0){8}}
\put(774,511){\line(-1,0){8}}
\put(782,510){\line(-1,0){8}}
\put(861,584){\line(-1,0){8}}
\put(632,369){\line(-1,0){8}}
\put(749,488){\line(-1,0){8}}
\put(580,312){\line(-1,0){8}}
\put(515,253){\line(-1,0){8}}
\put(852,591){\line(-1,0){8}}
\put(761,521){\line(-1,0){8}}
\put(948,680){\line(-1,0){8}}
\put(706,420){\line(-1,0){8}}
\put(941,673){\line(-1,0){8}}
\put(928,660){\line(-1,0){8}}
\put(543,284){\line(-1,0){8}}
\put(767,490){\line(-1,0){8}}
\put(677,446){\line(-1,0){8}}

\put(889,633){\line(1,0){8}}
\put(816,593){\line(1,0){8}}
\put(911,663){\line(1,0){8}}
\put(632,413){\line(1,0){8}}
\put(945,680){\line(1,0){8}}
\put(817,586){\line(1,0){8}}
\put(893,643){\line(1,0){8}}
\put(838,615){\line(1,0){8}}
\put(774,519){\line(1,0){8}}
\put(782,544){\line(1,0){8}}
\put(861,614){\line(1,0){8}}
\put(632,388){\line(1,0){8}}
\put(749,512){\line(1,0){8}}
\put(580,327){\line(1,0){8}}
\put(515,271){\line(1,0){8}}
\put(852,622){\line(1,0){8}}
\put(761,540){\line(1,0){8}}
\put(948,680){\line(1,0){8}}
\put(706,420){\line(1,0){8}}
\put(941,673){\line(1,0){8}}
\put(928,667){\line(1,0){8}}
\put(543,297){\line(1,0){8}}
\put(767,529){\line(1,0){8}}
\put(677,463){\line(1,0){8}}

\put(889,633){\line(-1,0){8}}
\put(816,593){\line(-1,0){8}}
\put(911,663){\line(-1,0){8}}
\put(632,413){\line(-1,0){8}}
\put(945,680){\line(-1,0){8}}
\put(817,586){\line(-1,0){8}}
\put(893,643){\line(-1,0){8}}
\put(838,615){\line(-1,0){8}}
\put(774,519){\line(-1,0){8}}
\put(782,544){\line(-1,0){8}}
\put(861,614){\line(-1,0){8}}
\put(632,388){\line(-1,0){8}}
\put(749,512){\line(-1,0){8}}
\put(580,327){\line(-1,0){8}}
\put(515,271){\line(-1,0){8}}
\put(852,622){\line(-1,0){8}}
\put(761,540){\line(-1,0){8}}
\put(948,680){\line(-1,0){8}}
\put(706,420){\line(-1,0){8}}
\put(941,673){\line(-1,0){8}}
\put(928,667){\line(-1,0){8}}
\put(543,297){\line(-1,0){8}}
\put(767,529){\line(-1,0){8}}
\put(677,463){\line(-1,0){8}}

\put(889,633){\line(0,1){4}}
\put(816,593){\line(0,1){4}}
\put(911,663){\line(0,1){4}}
\put(632,413){\line(0,1){4}}
\put(945,680){\line(0,1){4}}
\put(817,586){\line(0,1){4}}
\put(893,643){\line(0,1){4}}
\put(838,615){\line(0,1){4}}
\put(774,519){\line(0,1){4}}
\put(782,544){\line(0,1){4}}
\put(861,614){\line(0,1){4}}
\put(632,388){\line(0,1){4}}
\put(749,512){\line(0,1){4}}
\put(580,327){\line(0,1){4}}
\put(515,271){\line(0,1){4}}
\put(852,622){\line(0,1){4}}
\put(761,540){\line(0,1){4}}
\put(948,680){\line(0,1){4}}
\put(706,420){\line(0,1){4}}
\put(941,673){\line(0,1){4}}
\put(928,667){\line(0,1){4}}
\put(543,297){\line(0,1){4}}
\put(767,529){\line(0,1){4}}
\put(677,463){\line(0,1){4}}

\put(889,633){\line(0,-1){4}}
\put(816,566){\line(0,-1){4}}
\put(911,657){\line(0,-1){4}}
\put(632,366){\line(0,-1){4}}
\put(945,680){\line(0,-1){4}}
\put(817,573){\line(0,-1){4}}
\put(893,604){\line(0,-1){4}}
\put(838,603){\line(0,-1){4}}
\put(774,511){\line(0,-1){4}}
\put(782,510){\line(0,-1){4}}
\put(861,584){\line(0,-1){4}}
\put(632,369){\line(0,-1){4}}
\put(749,488){\line(0,-1){4}}
\put(580,312){\line(0,-1){4}}
\put(515,253){\line(0,-1){4}}
\put(852,591){\line(0,-1){4}}
\put(761,521){\line(0,-1){4}}
\put(948,680){\line(0,-1){4}}
\put(706,420){\line(0,-1){4}}
\put(941,673){\line(0,-1){4}}
\put(928,660){\line(0,-1){4}}
\put(543,284){\line(0,-1){4}}
\put(767,490){\line(0,-1){4}}
\put(677,446){\line(0,-1){4}}

\sbox{\plotpoint}{\rule[-0.400pt]{0.800pt}{0.800pt}}%
\put(350,82){\usebox{\plotpoint}}
\put(352,82){\line(1,1){600}}

\put(677,446){\line(0,1){17}}
\put(767,490){\line(0,1){39}}
\put(543,284){\line(0,1){13}}
\put(706,420){\line(0,1){1}}
\put(928,660){\line(0,1){7}}
\put(941,673){\line(0,1){1}}
\put(948,680){\line(0,1){1}}
\put(761,521){\line(0,1){19}}
\put(852,591){\line(0,1){31}}
\put(515,253){\line(0,1){18}}
\put(580,312){\line(0,1){15}}
\put(749,488){\line(0,1){24}}
\put(632,369){\line(0,1){19}}
\put(861,584){\line(0,1){30}}
\put(782,510){\line(0,1){34}}
\put(774,511){\line(0,1){8}}
\put(838,603){\line(0,1){12}}
\put(893,604){\line(0,1){39}}
\put(817,573){\line(0,1){13}}
\put(945,680){\line(0,1){1}}
\put(632,366){\line(0,1){47}}
\put(911,657){\line(0,1){6}}
\put(816,566){\line(0,1){27}}
\put(889,633){\line(0,1){1}}

\sbox{\plotpoint}{\rule[-0.200pt]{0.400pt}{0.400pt}}%
\put(350.0,82.0){\rule[-0.200pt]{0.400pt}{144.058pt}}
\put(350.0,82.0){\rule[-0.200pt]{144.058pt}{0.400pt}}
\end{picture}

\end{center}
\caption{Comparing the $L^1$ boxcar van Rossum metric with the $p=1$
  and $p=10$ ${\mathcal L}_p$ metrics. Here $\tilde{h}$ values for the
  $p=1$ and $p=10$ metrics are graphed against the values for the
  $L^1$ van Rossum metric with a boxcar filter with optimal values of
  the $q$ or $\tau$ parameter has been used for each site. The $p=1$
  and $p=10$ values are marked by a horizontal line, the two values
  for a given site are joined by a vertical line. Since increasing $p$
  never decreases $\tilde{h}$ between $p=1$ and $p=10$, the topmost of
  the two horizontal lines corresponds to $p=10$. The \lq{}x=y\rq{}
  line is also plotted for clarity.\label{Compare}}
\end{figure}
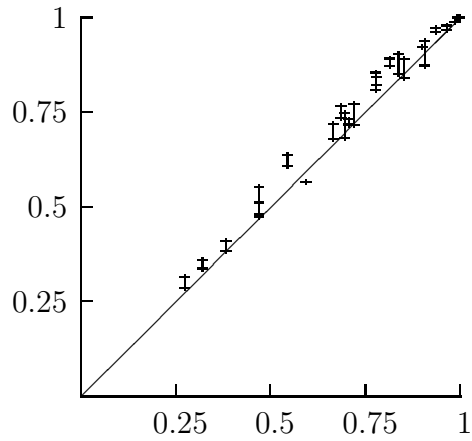

The pairing of near coincident spikes is the key distinction between
the Victor-Purpura metrics and the van Rossum metrics, a family of
spike train metrics which are calculated by comparing reconstructed
rate functions. In fact, computationally, an $L^1$ van Rossum metric
with a boxcar filter measures a very similar distance to the
Victor-Purpura metric \cite{HoughtonSen2006a} and, as seen in
Fig.~\ref{Compare}, has a very similar performance. The way in which
${\mathcal L}_p$ Victor-Purpura metric performance increased with $p$
seems to indicate that the significant temporal structure of spikes is
not fully accounted for by the temporal structure of a spike rate.

\section*{Acknowledgments}
C.H. is supported by Science Foundation Ireland grant
08/RFP/MTH1280. He thanks Kamal Sen for the use of the zebra finch
data analyzed here and Jonathan Victor for comments on an early draft
of this comment.

\end{document}